\documentclass[a4paper]{article}

\usepackage[english]{babel}
\usepackage[utf8x]{inputenc}
\usepackage[T1]{fontenc}

\usepackage[a4paper,top=3cm,bottom=2cm,left=3cm,right=3cm,marginparwidth=1.75cm]{geometry}

\usepackage{amsmath}
\usepackage{graphicx}
\usepackage[colorinlistoftodos]{todonotes}
\usepackage[colorlinks=true, allcolors=blue]{hyperref}
\usepackage{authblk}

\title{Sources of Variance in Two-Photon Microscopy Neuroimaging}
\author{Kyongche Kang}
\author{Jinsub Hong}
\author{Hannah Worrall}
\affil{Department of Statistics\\Carnegie Mellon University}

\begin{document}
\date{May 2013}
\maketitle

\begin{abstract}
 Two-photon laser scanning microscopy is widely used in a quickly growing field of neuroscience. It is a fluorescence imaging technique that allows imaging of living tissue up to a very high depth to study inherent brain structure and circuitry. Our project deals with examining images from two-photon calcium imaging, a brain-imaging technique that allows for study of neuronal activity in hundreds of neurons and and. As statisticians, we worked to apply various methods to better understand the sources of variations that are inherent in neuroimages from this imaging technique that are not part of the controlled experiment. Thus, images can be made available for studying the effects of physical stimulation on the working brain. Currently there is no system to examine and prepare such brain images. Thus we worked to develop methods to work towards this end. Our data set had images of a rat’s brain in two states. In the first state the rat is sedated and merely observed and in the other it is repeatedly simulated via electric shocks. We first started by controlling for the movement of the brain to more accurately observe the physical characteristics of the brain. We analyzed how the variance of the brain images varied between pre and post stimulus by applying Levene’s Test. Furthermore, we were able to measure how much the images were shifted to see the overall change in movement of the brain due to electrical stimulus. Therefore, we were able to visually observe how the brain structure and variance change due to stimulus effects in rat brains.\end{abstract}

\section{Introduction}

With development of neuroscience, researchers are actively engaging in studying inner workings of brains. One popular method is to take images directly by two-photon microscopy, a fluorescence imaging technique that allows imaging of living tissue up to a very high depth, up to about one millimeter. This allows scanning of the emission of two different fluorescents simultaneously, thus highlighting different parts of the brain. The issue with images obtained through this process is that there are various sources of variance across the images that make it difficult to isolate the variations of brain structures by conditions imposed on the subject. In this analysis, we identify sources of variations across the images and explore ways to control for them.

\section{Exploratory Data Analysis}

\subsection{Design of Experiment}

The researcher recorded the videos of a rat’s brain in two different states - resting and stimulated states. The subject rat is sedated and strapped, and brain activity was recorded by two-photon microscopy on a small portion of the rat’s brain. To keep the subject rat alive, it was placed on a respirator. These are experiments recorded sequentially, first resting, then stimulus, in the same part of the brain of the same rat. The stimulated state was reached by periodically applying an electric shock to the rat’s forepaw. There are 20 trials of stimulation - at each of the trial, the rat is introduced to 12, 1.5mA electric shocks. Each shock is 1ms long with a gap of 167ms between shocks. Each trial lasts about 2 seconds. \\ 
\\ To highlight the cells of the brain, two fluorescent dyes used.  The structural dye is called Sulforhodamine 101, and it marks astrocytes but does not change its properties over time. The functional dye is called Oregon Green Bapta, and it marks neurons and astrocytes. The former is highlighted with red, and the latter is highlighted with green color.

\begin{figure}
\centering
\includegraphics[width=1.0\textwidth]{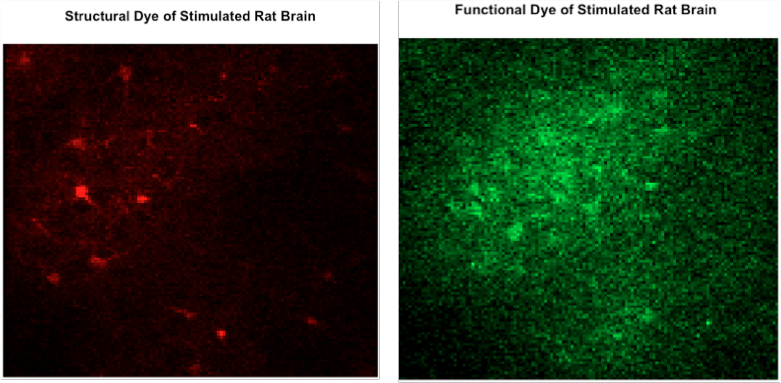}
\caption{\label{fig:fig1}Images of Structural and Functional Dye of a rat in stimulated state}
\end{figure}

\subsection{Data}
Our data consists of two sets images of a rat’s brain in a resting state and in a stimulated state. They are taken from the videos recorded by two-photon microscopy. Each image is taken at one-eighth of a second from the video. Each data set is a four-dimensional array. First dimension indicates the type of dye used, structural and functional. Second dimension represents time point when each image was taken. In resting state data, there are 2400 images, which translates into 300 seconds long sequence of images. In stimulated state data, there are 3880 images, which translates into 485 seconds long sequence of images. Third and fourth dimensions represent rows and columns of images. 
\\
\\
Our data also has information on the rat’s biological activity. The biopac element of the experiment gathered information on many variables such as heart rate, blood pressure, respiration rate, ECG, as well as time stamps for when each stimulation trial began. All of this information was collected at a rate of 1000 samples per second. As the rat was on a respirator, sedated, and strapped, most of its vital signs were actively regulated by the experiment.
\\
\\ 
Our data are composed entirely of images. When we were examining them we found that when we took the mean of each pixel over time and plotted it, it becomes easier to visualize position of cells in the brain, especially in the functional dye. We also see that it is possible to line up the functional and structural dyes and pick out which cells were astrocytes in the functional dye. In order to make this clear we overlayed the structural dye on the functional dye. Figure \ref{fig:fig1} shows the images of two dyes overlayed. The highlighted cells are very likely astrocytes.

\begin{figure}
\centering
\includegraphics[width=0.7\textwidth]{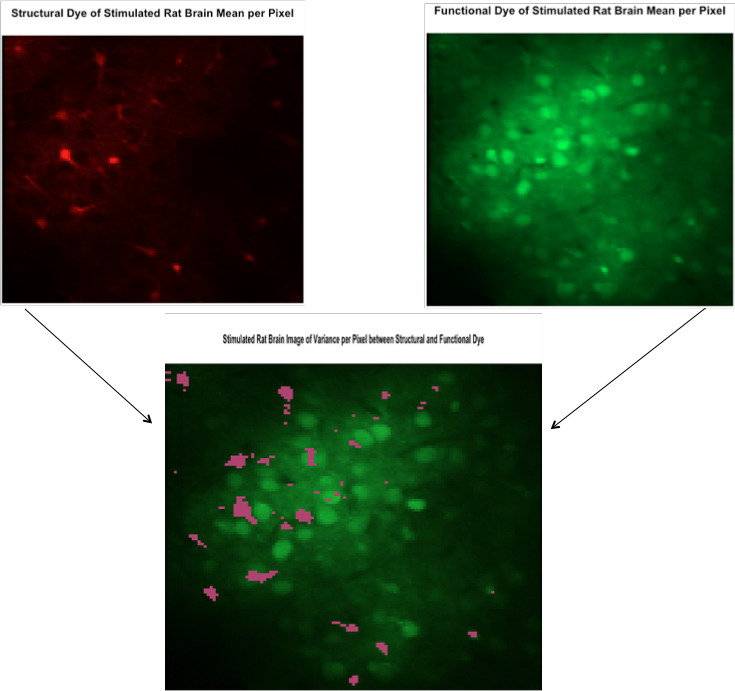}
\caption{\label{fig:fig2}Comparison of functional and structural dye of neuroimages}
\end{figure}

\section{Controlling for Brain Movement}

No parts of living beings are stationary. When the images of brain were taken, although the subject rat was strapped and images were taken at exactly the same part of the brain, the brain itself was not stationary. In fact, brain and their cells were actively moving. This poses a difficulty in analyzing and comparing across images, because when we compare the structure of the brain, we see change in intensities of certain parts of the brain, and this may be confounded by the fact that the changes were merely caused by the same cells’ movements. Also, any statistical test for analyzing the difference in variance of image pixels won’t be able to isolate the change in variation due to cells merely moving in different directions. 
Thus, we identify this positional variation introduced during brain movement as one of the important source of variations in neuroimaging.

\subsection{Mechanisms}

The main idea behind image alignment is that when we overlay images on top of each other, we would want to see all the spots of intensities and structures match across the images. 
We approach this problem by utilizing a transformation function that shifts and rotates images. We choose a reference image in the middle of the experiment to spatially target images to align with the reference image. We optimize the image manipulation function by minimizing the least squared errors of pixels with reference image with transformation function. 

\subsection{Post-Transformation Images}

We compare the change in variance of  prior and post image alignment. Figure 3 shows the scatterplot of logged mean against logged variance of each pixel of images. We see that after the alignment, variance over mean seems to be coming down, which is evident by observations coming closer to the straight line. After aligning the images, the variance of pixels decreased by 12.5\%. 
\\
\\
Figure \ref{fig:fig4} illustrates the movement of the brain. Overlaying unaligned and aligned images together, we see that the highlighted orange color represents unaligned parts of the brain. We clearly see that there are brain movements in the images.

\begin{figure}
\centering
\includegraphics[width=1.1\textwidth]{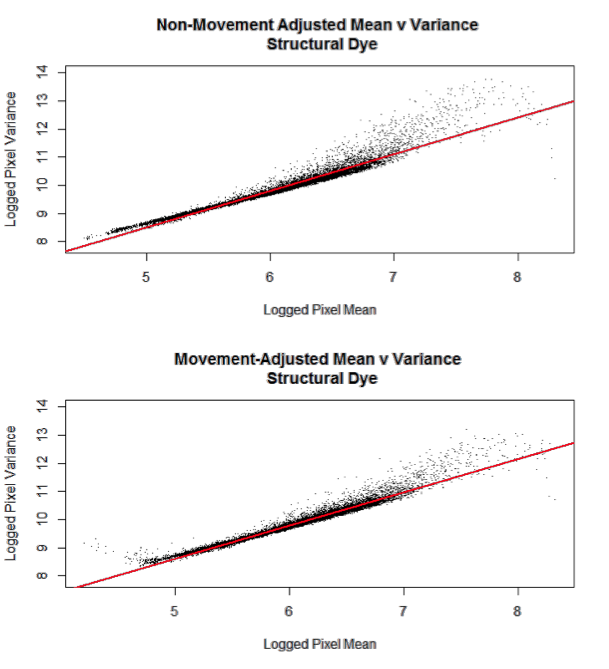}
\caption{\label{fig:fig3}Mean versus Variance of Pixels plots before and after image alignment}
\end{figure}

\begin{figure}
\centering
\includegraphics[width=1.1\textwidth]{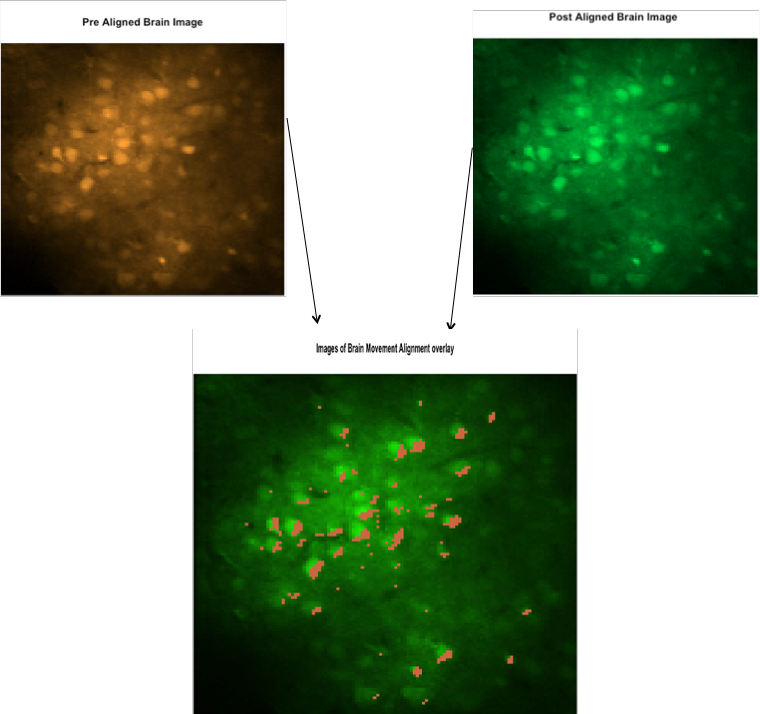}
\caption{\label{fig:fig4}Image of Functional Dye before and after image alignment}
\end{figure}

\section{Stimulus Effect on Brain Images}

\begin{figure}
\centering
\includegraphics[width=1.0\textwidth]{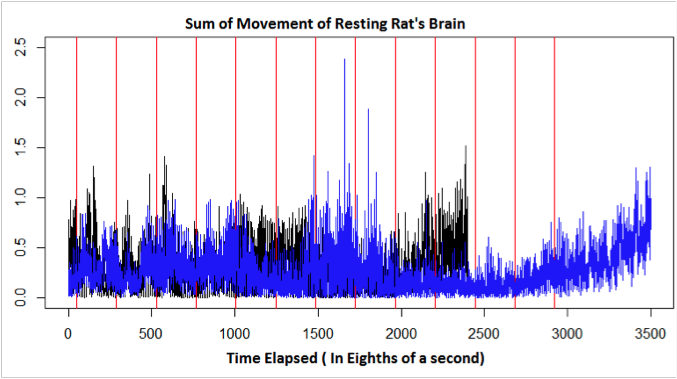}
\caption{\label{fig:fig5}Time-series plot of sum of movement of resting rat’s brain}
\end{figure}

We sought to test whether or not the stimulus affected the physical movement of the brain throughout the experimental trials. If the real goal is to visualize and or observe the firing of neural structure, potential movement in the brain due to the stimulus may obscure images and produce undesirable results. Thus, knowing that as many variables as physically possible were held constant throughout the experiment (a respirator, strappings on the rat, evenly time experimental trials, and sedation of the rat) we looked to test whether or not the overall variance structure of the rat’s brain movement over time differed between the stimulated and unstimulated trials.
\\
\\
Our first step was to plot the time series of the total sum of brain movement (whether natural or not) over times for both the stimulated and unstimulated states. Overlaying the two time series with red bars indicating each instant of a stimulus reveals no obvious visual clues as to whether or not there was a difference between the stimulated and unstimulated brain movement, as shown in Figure \ref{fig:fig5} (Blue time series represents the stimulated rat, whereas the black represents the resting state rate)
\\
\\
However, to more rigorously test whether or not there was a clear-cut difference between brain movement between the stimulated and unstimulated states, we compared the underlying variance structures of the rat in its two different states.

\subsection{Levene’s Test}

\begin{center}
\includegraphics[width=0.5\textwidth]{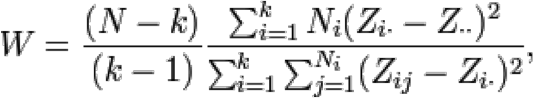}
\end{center}

Levene’s test is a frequently used, non-parametric test used to test whether two samples from a population exhibited the same underlying variance structure. It is most commonly used as a precursor for parametric modeling. For our purposes however, we were simply interested in whether or not the underlying structures were different for brain movement over time and for heart rate over time. Should the two structures be different between stimulated and unstimulated state, we can tentatively conclude that the stimulus did affect the rat over time.

\subsection{Application: Brain Movement}
We first sought to address whether or not the stimulus affected the movement of the brain over time. This is important, as anyone with intention of visually observing the effect of stimulus on the actual structure of the brain needs to be confident that the brain is not moving due to indirect effects of the stimulus. After applying our non-parametric Levene’s Test, we were able to acquire an estimate of 0.29, insignificant, and therefore we tentatively concluded that the stimulus did not cause the brain itself to shift dramatically enough for the actual image of the brain to move.

\subsection{Application: Heart Rate}
Another possible source of image variation we explored was the effect of electric stimulus on biological factors of the rat. The rat was on a respirator, strapped, and sedated throughout these experiments and therefore, only left heart rate as a strong potential indirect variance source to be investigated.

\begin{figure}[!h]
\includegraphics[width=0.5\textwidth]{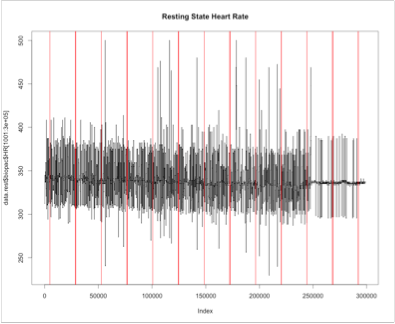}
\includegraphics[width=0.5\textwidth]{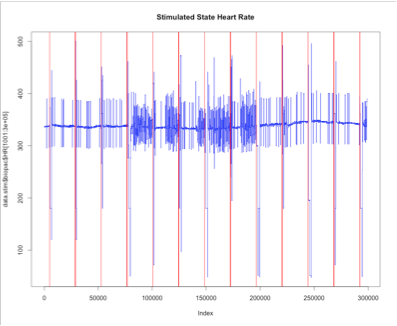}
\caption{\label{fig:fig6}Comparison of times-series plots of heart rate in resting and stimulated states}
\end{figure}

The red lines indicate every electric shock on the stimulated rat (the blue wave) whilst the red lines indicate where the shocks would have been on the unstimulated rat (the blue wave). After applying our non-parametric Levene’s Test, we obtained a significant p-value of 0.0002 which lead us to tentatively conclude that the rat’s heart was affected by the stimulus. However, the change in heart rate did not lead to any change in brain movement. This lead us to wonder what would happen had the rat not been on a respirator. Would other affected biological factors end up causing noise in the brain image data?

\section{Neuron Intensity}

We were interested in whether or not there was an observable difference in the neurons of our rat when it was resting and when it was being stimulated. In other words, does the stimulus cause neurons to fire more frequently? Supposedly if this were true then it would be possible to pick out the difference in our data. When a neuron fires it becomes brighter. Therefore, if a neuron is firing more frequently, its mean intensity would be higher than its mean intensity if it were not. Therefore, we are looking for groups of pixels which likely represent neurons that have a different mean value in the resting state than the stimulated state. Before we start looking for such pixels, however, we need to clean the data some more. The average intensity of the images changes over time. Therefore, if we want to look at the mean intensity it would be a good idea to make the average intensity of each image the same. We did this through mean equalization.
\\
We were also interested in reducing the variance in our data. We hoped that using mean equalization would reduce the variance, and we were interested in how much it would do so.

\subsection{Mean Equalization}

The goal of mean equalization is to make the average intensity of each image the same. This requires the choice of a standard intensity. We chose the mean intensity of the mean intensities. In other words, we found the mean intensity of each image at each time point. We then found the mean of these mean intensities, and used it as our standard. We then multiplied each pixel at each time point by the standard divided by the mean of the image that pixel came from. This was then the mean-equalized intensity for that pixel. We did this for the resting images and the simulated images and used the resulting data for the rest of our analysis. We only did this for the functional dye, as it was the only dye that showed neurons.

\subsection{Results}

\begin{figure}[!h]
\centering
\includegraphics[width=0.6\textwidth]{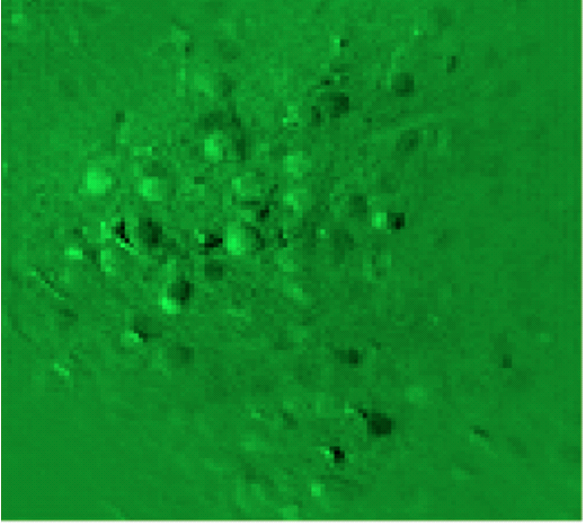}
\caption{\label{fig:fig7}Differences Mean Equalized images of Stimulated and Resting Rat’s Brain Image post movement adjustment and mean equalization}
\end{figure}

After applying mean equalization to our data we found the mean intensity over time for each pixel for both the resting rat and the simulated rat. After doing so we plotted the resulting images for the calcium dye. We then subtracted the resting rat image from the stimulated rat image. The difference in mean equalized images of stimulated vs. resting rat's brain image is in Figure \ref{fig:fig7} 

\noindent By examining the difference, we can clearly see the shape of cells that are activated during stimulation. Therefore, there is a significant difference in intensity between the stimulated rat and resting rat. This indicates that the stimulus causes some neurons to fire more often than they do while the rat is in a resting state.  
We also calculated the total variance of the images before and after mean equalization was applied. We found that mean equalization decreased variance in the resting rat by .6\%, and decreased variance in the stimulated rat by 1.3\%. This further indicates that the stimulation caused visible effects in the brain. Mean equalization makes every image’s mean the same. If doing so reduces variance more in the stimulated rat, then there were time frames when the intensity was more different from the mean than in the resting rat. Therefore, it is plausible that this increased deviation was due to the stimulus.

\section{Conclusion}

We identified sources of variances in two-photon microscopy neuroimaging of a rat. We saw that the adjusting for brain movement and effects of stimulus through mean equalization reduced variance significantly.. Further, taking the difference of mean equalized stimulated and resting states images revealed evidence that there seems to be a difference in intensity between them. Through Levene’s test we found that the heart rate of the rat in resting and stimulated states were different, however the change in heart rate did not seem to change the movement of the brain. By analyzing the variance of the brain images varied between pre and post stimulus by applying Levene’s Test, we were able to identify overall change in movement of the brain due to electrical stimulus. This enables us to visually observe the brain structure and activation of neurons.Mapping of the brain has been largely a mystery for humankind. With application in our research, stimulus can help us identify the various parts of the brain that are responsible for functions of our body.

\end{document}